\documentclass{emulateapj}

\def\deg{$^{\circ} $}

\newcommand{\kms}{km~s$^{-1}$}
\newcommand{\kmsm}{km~s$^{-1}$~Mpc$^{-1}$}
\newcommand{\Ha}{$\rm H\alpha$}

\newcommand{\hi}{{H{\sc i}}}

\newcommand{\rband}{{\em r}-band}
\newcommand{\iband}{{\em I}-band}

\newcommand{\whi}{$W_{50}$}
\newcommand{\whicor}{$W_{50}^c$}
\newcommand{\x}{$\times$}
\newcommand{\about}{$\sim$}
\newcommand{\Msun}{M$_\odot$}
\newcommand{\Mhi}{$M_{\rm HI}$}
\newcommand{\MhiLi}{$M_{\rm HI}/L_{\rm I}$}
\newcommand{\Mstar}{$M_{\rm star}$}

\newcommand{\Mi}{$M_{\rm I}$}
\newcommand{\Li}{$L_{\rm I}$}
\newcommand{\Lsun}{L$_\odot$}

\slugcomment{Accepted for publication in ApJ Letters}
\shorttitle{A Pilot Survey of HI in Field Galaxies at z\about 0.2}
\shortauthors{Catinella et al.}

\begin{document}

\title{A Pilot Survey of HI in Field Galaxies at Redshift z\about 0.2}

\author{Barbara Catinella\altaffilmark{1,2}, Martha P. Haynes\altaffilmark{3}, 
Riccardo Giovanelli\altaffilmark{3}, Jeffrey P. Gardner\altaffilmark{4}, \& Andrew J. Connolly\altaffilmark{5}}

\altaffiltext{1}{Max-Planck-Institut f\"{u}r Astrophysik, D-85741
Garching, Germany; bcatinel@mpa-garching.mpg.de.}
\altaffiltext{2}{Arecibo Observatory, HC3 Box 53995, Arecibo, PR 00612, USA.
The Arecibo Observatory is part of the National Astronomy and
Ionosphere Center, which is operated by Cornell University under a
cooperative agreement with the National Science Foundation.}
\altaffiltext{3}{Center for Radiophysics and Space Research and
  National Astronomy and Ionosphere Center, Cornell University,
  Ithaca, NY 14853, USA.}
\altaffiltext{4}{Department of Physics, University of Washington,
  Seattle, WA 98195, USA.}
\altaffiltext{5}{Department of Astronomy, University of Washington,
  Seattle, WA 98195, USA.}

\begin{abstract}

We present the first results of a targeted survey carried out
with the 305~m Arecibo telescope to detect \hi-line emission from
galaxies at redshift $z>0.16$. The targets, selected from the 
Sloan Digital Sky Survey database, are non-interacting disk galaxies
in relatively isolated fields. We present here the \hi\ spectra and
derived \hi\ parameters for ten
objects detected in this pilot program. All are massive disk galaxies 
in the redshift interval 0.17$-$0.25 (i.e. 2$-$3 Gyr look-back time), 
with \hi\ masses \Mhi~$=3-8$\x 10$^{10}$ \Msun\ and high gas mass
fractions (\hi - to - stellar mass ratios \about 10$-$30\%).
Our results demonstrate the efficacy of exploiting Arecibo's
large collecting area to measure the \hi\ mass and
rotational velocity of galaxies above redshift $z=0.2$. In particular, 
this sample includes the highest redshift detections
of \hi\ emission from individual galaxies made to date.
Extension of this pilot program will allow us to study the \hi\ 
properties of {\em field} galaxies at cosmological distances, thus 
complementing ongoing radio synthesis observations of cluster samples at
$z$\about 0.2. 

\end{abstract}


%-------------------------------------------------------------------------------
\section{Introduction}\label{s_intr}
%-------------------------------------------------------------------------------
%
Because cold atomic hydrogen represents the reservoir for future 
star formation and the physics of the 21 cm transition is relatively simple, 
detection of the \hi\ line emission has the potential to contribute
a key quantitative ingredient to models of how galaxies 
acquire and retain gas and convert it into stars. Of critical importance
is the tracking of the gas content over the redshift interval between
$z \sim 1$ and 0, i.e. over the period when the observed cosmic star 
formation rate density has declined by an order of magnitude 
\citep[e.g.,][and references therein]{lilly96,madau96,bell05}.
However, our knowledge of the \hi\ 21~cm line emission properties 
of galaxies is currently limited to the very local universe, and
only recently have technical improvements in radio frequency
receivers and back-ends made attempts to explore the \hi\ emission
from galaxies at $z> 0.1$ in any practical sense.

The first detection of \hi\ emission from a galaxy at cosmological
distance ($z=0.1766$) was made by \citet{zvv01} with the Westerbork
Synthesis Radio Telescope (WSRT). More recently, \cite{ver07}
successfully carried out a pilot program with the WSRT to image 
$z\sim 0.2$ galaxy clusters at 21 cm. They targeted the
Abell 963 ($z=0.206$) and Abell 2192 ($z=0.188$) clusters for 20\x
12 hours and 15\x 12 hours respectively, and detected \hi\
emission from 42 galaxies. The Giant Metrewave Radio Telescope 
was used by \citet{lah07} to constrain the \hi\ content of star
forming galaxies at $z=0.24$. They obtained an average \hi\ spectrum
by co-adding the non-detections of 121 galaxies with known optical
redshifts. However, their detection is at best marginal.

Historically, Arecibo's huge collecting area has
offered a critical advantage to \hi\ line studies of extragalactic
objects, because sensitivity is of paramount importance to the detection
of weak \hi\ signals. Recent improvements to the
Arecibo instrumentation, both front-end and back-end, are allowing the
practical exploration of the frequency range corresponding to \hi\ at
$z>0.1$. Most notably, the installation of the new, single-pixel
L-band receiver (``L-wide'') in February 2003 has given access
to frequencies down to 1120 MHz (corresponding to redshifted \hi\
emission at $z=0.27$) with improved system noise and performance.
Naturally, single-dish \hi\ observations of intermediate redshift 
galaxies are challenged by a lack of angular resolution. 
As the density of galaxies per unit solid angle increases with redshift, 
so also does the probability of finding multiple sources within
the telescope beam. However, beam confusion can be broken
kinematically, if high-quality optical redshifts are available for the
sources within the beam. The availability of imaging and spectroscopic
data from the Sloan Digital Sky Survey \citep[SDSS;][]{yor00} has enabled
the identification of sources suitable for long-integration
\hi\ line observations with the Arecibo telescope.
In this Letter, we report the first results of a pilot survey
carried out at Arecibo to detect \hi\ emission from disk galaxies at
$z\sim 0.2$. 

All the distance-dependent quantities in this work are calculated
assuming $H_0 = 71$ \kmsm, $\Omega_{\rm M}=0.27$ and 
$\Omega_{\rm vac}=0.73$.

\vspace{24pt}

%-------------------------------------------------------------------------------
\section{Sample, Observations and Data Reduction}\label{s_survey}
%-------------------------------------------------------------------------------

The initial goal of this survey was to explore the feasibility of detecting
\hi\ line emission from galaxies beyond the redshift range of past 
Arecibo surveys, that is at redshifts $z >$0.05. 
First observations successfully detected over 50
galaxies with $0.05\leq z\leq 0.09$, and the program was extended to
higher redshifts with the availability of the new L-wide receiver.
In the end, the frequency interval corresponding to 
$0.11 < z < 0.16$ was excluded because of the particularly hostile 
Radio Frequency Interference (RFI) environment. Here, we report
results only for objects with $z > 0.16$; the full dataset will be
presented elsewhere (B. Catinella et al., in preparation).

The targets for \hi\ spectroscopy were extracted from the SDSS
database according to the following criteria: 
(a) observable from Arecibo during night-time (to minimize RFI); 
(b) redshift $0.16 < z < 0.27$ and not corresponding to the
frequency of known RFI within that interval. This redshift range
corresponds to the 1120$-$1220 MHz frequency interval and is dictated by
the instrumentation (see below); 
(c) inclination $i \geq 45$\deg\ (for use in disk scaling relations);
(d) presence of \Ha\ emission with line width between 100 and 700
\kms\ and equivalent width larger than 5 \AA;
(e) exponential disk profile;
(f) non-interacting, undisturbed, disk appearance in the SDSS images.
The SDSS images of the candidate targets were carefully inspected
to exclude objects with interacting or peculiar appearance, and/or
with \hi\ emission possibly contaminated by that of other, nearby
objects lying within the Arecibo beam. We also discarded targets lying
in the vicinity of continuum sources that would cause 
ripples in the baselines.

The \hi\ data for the $z > 0.16$ sample were collected
during four observing runs, distributed between Fall 2003 and  Spring
2007. Including all the overheads, the total time spent on this part
of the survey was 280 hours. The observations were conducted 
in standard {\em position-switching} mode: each
observation consisted of an {\it on/off} source pair, each integrated
for 4 minutes (5 minutes in the last observing run), followed by the
firing of a calibration noise diode. The {\it off}
observation of a ``blank sky'' region was performed by tracking
exactly the same range of azimuth and zenith angle as was previously
occupied by the {\it on} source, as commonly done at Arecibo 
to optimize the bandpass subtraction.
The {\em on-source} integration time per object varied between 80
minutes and 4.7 hours, primarily depending on the redshift and on the
observed rotational velocity of the target (for the same \hi\ flux,
slower rotators are easier to detect).
The spectra were acquired using the L-band wide receiver, which operates in
the frequency range 1120$-$1730 MHz. We also used a 1120$-$1220 MHz
filter and a 750 MHz narrow band front-end filter to limit the impact
of RFI on our observations. Although sensitivity
requirements suggest long accumulations of data, RFI considerations
dictate fast spectral dumping, because much RFI is of short
duration. Thus spectra were recorded every second with 9-level
sampling. Two correlator boards, each configured for 12.5 MHz
bandwidth, one polarization, and 2048 channels per spectrum
(yielding a velocity resolution of 1.8 \kms\ at 1200 MHz before
smoothing) were centered at or near the frequency
corresponding to the SDSS redshift of the target. Minimizing the
number of different board settings facilitates the identification of
RFI. In order to better discriminate terrestrial sources
from cosmic ones, we also chose not to apply any on-line Doppler
correction for the motion of the Earth (the correction is applied 
during off-line processing).   

The data reduction was performed using our own routines based on the
standard Arecibo data processing library and will be described in full detail elsewhere. 
In summary, the data processing of each polarization and
{\it on-off} pair includes the following steps: Hanning smoothing, 
bandpass subtraction, RFI excision, and flux calibration. A
final spectrum for each of the two orthogonal linear polarizations
for each galaxy is constructed by the combination
of good quality records (those without serious RFI or standing waves),
weighting each by a factor $1/rms^2$, where $rms$ is the root mean square
noise measured in the signal-free portion of the spectrum. 
The two polarizations are then averaged, yielding the
final spectrum. After boxcar smoothing and baseline subtraction, the \hi-line
profiles are ready for the measurement of redshift, rotational
velocity and integrated  \hi\ line flux. Our measurement
technique is explained, e.g., in \citet[][\S 2.2]{widths}.

%-------------------------------------------------------------------------------
\section{Results}\label{s_res}
%-------------------------------------------------------------------------------

The total number of galaxies targeted by our $z>0.16$ pilot survey is
41 (this number does not include several other objects that were
initially targeted but had to be abandoned after short integrations
because of RFI or other problems). Half of the targeted galaxies are sure
detections with high-quality \hi\ profiles, a subset of which is
presented in this Letter. Of the remaining half, only \about 1/3 are
clearly non-detections; the remaining objects include marginal
detections, uncertain profiles or otherwise problematic cases that are
still being analyzed.

\begin{figure}
\epsscale{1.1}
\plotone{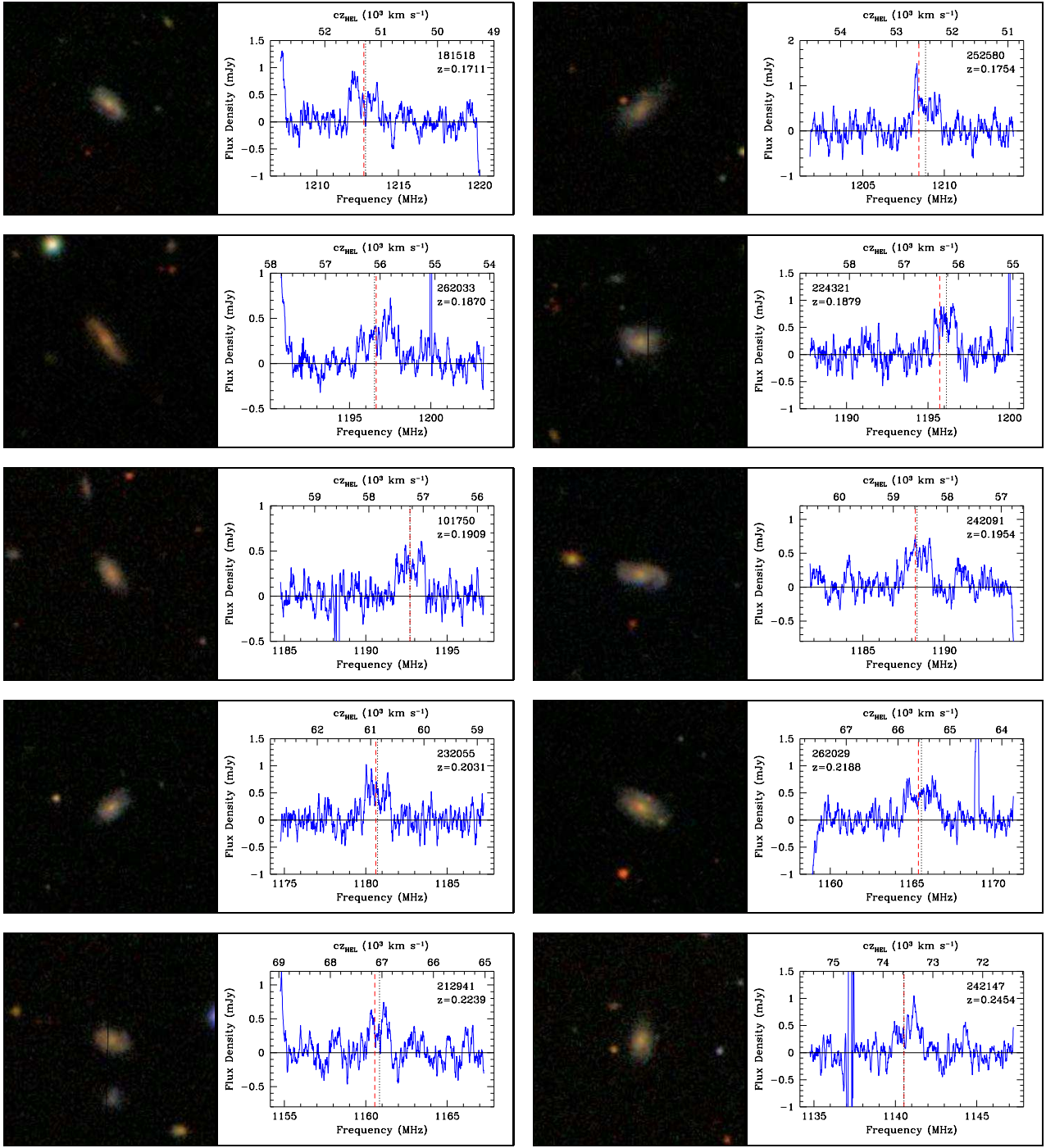}
\caption{SDSS postage stamp images and \hi\ spectra of the 10
Arecibo detections discussed in this work. Each SDSS image is 1 arcmin
square. The \hi\ spectra are calibrated, smoothed and
baseline-subtracted. Each spectrum is labeled with its AGC
identifier and SDSS redshift. The central heliocentric velocity
(frequency) corresponding to the SDSS redshift and that measured from
the spectrum are indicated with a dashed and a dotted line,
respectively. The {\em rms} noise per channel at the final velocity
resolution (27 \kms\ at 1200 MHz), measured in the RFI-free portions
of the baseline, varies between 0.1 and 0.2 mJy for these spectra.
}\label{plate}
\end{figure}

Figure~\ref{plate} (Plate 1) presents 10 of the best quality \hi\ detections
of $z$\about 0.2 galaxies obtained at Arecibo, ordered by increasing
redshift (as indicated next to each spectrum). These galaxies are not
necessarily the highest signal-to-noise detections, but have been
chosen to span the entire $0.16<z<0.25$ redshift interval covered by
our pilot program. Each panel of Fig.~\ref{plate} shows the SDSS
image (1\arcmin\ size, which is equivalent to about 1/4 of the half
power full width of the Arecibo beam at 1200 MHz) and the \hi\
spectrum of the targeted galaxy (the full 12.5 MHz bandwidth is
shown). The \hi-line profiles are boxcar
smoothed by 15 channels, corresponding to a velocity resolution
of 27 \kms\ at 1200 MHz, and
baseline-subtracted. The spectra are labeled with two numbers, the
Arecibo General Catalog (AGC; a private database maintained at Cornell
University by MPH and RG) identifier and the SDSS redshift of the
galaxy. The frequency and heliocentric velocity (bottom and top $x$
axis, respectively) corresponding to the SDSS redshift are indicated
by a dashed line. The dotted line is at the redshift
measured from the \hi\ spectrum. The last three objects,
with $z>0.21$, represent the highest redshift detections of \hi-line
emission from individual galaxies made to date.

Table~\ref{table} lists SDSS and \hi\ parameters for the galaxies
presented in Fig.~\ref{plate}, similarly ordered by increasing
redshift.  The quantities are listed as follows: 
Columns (1) and (2) are the source identifiers in the AGC and SDSS databases. 
Column (3) is the SDSS redshift, $z_{\rm SDSS}$; the associated
uncertainty on the last digit is shown within brackets.
Column (4) is the absolute \iband\ magnitude in
the Cousins system, \Mi\ (see \S~\ref{ss_res}). 
Column (5) is the galaxy inclination to the line-of-sight in degrees,
$i$, estimated from SDSS tabulated axis ratios in \rband. 
Column (6) is the on-source integration time of the Arecibo
observation, $T_{\rm on}$, in minutes (only {\it on scans} that were
actually combined are counted). 
Column (7) is the redshift, $z$, measured from the \hi\ spectrum.
The error on the corresponding heliocentric velocity, $cz$, 
is half the error on the width, tabulated in the following column.
Column (8) is the observed velocity width of the source line profile
in \kms, \whi\ (measured at the 50\% of the peak levels), with its
statistical uncertainty.
Column (9) is the velocity width corrected for instrumental broadening
and cosmological redshift only, \whicor. No inclination or turbulent motion
corrections are applied.
Column (10) is the observed, integrated \hi-line flux density in Jy \kms,
$F \equiv \int S~dv$, measured on the smoothed and baseline-subtracted
spectrum. The statistical errors are calculated according to equation 2 of 
\citet[hereafter S05]{s05}.
Column (11) is the signal-to-noise ratio of the \hi\ spectrum, S/N,
estimated following \citet{saintonge07} and adapted to the velocity
resolution of our spectra. This definition of S/N accounts for the
fact that for the same peak flux a broader spectrum has more signal.
Column (12) is the base-10 logarithm of the \hi\ mass, \Mhi, in solar
units. The \hi\ masses of these galaxies vary between 
3 and 8\x 10$^{10}$ \Msun\ and were computed via: 
\begin{equation}
    \frac{M_{\rm HI}}{\rm M_{\odot}} = \frac{2.356\times 10^5}{1+z}
    \left[ \frac{d_{\rm L}(z)}{\rm Mpc}\right]^2
    \left(\frac{\int S~dv}{\rm Jy~km~s^{-1}} \right)
\label{eq_MHI}
\end{equation}
\noindent
where $d_{\rm L}(z)$ is the luminosity distance to the galaxy at
redshift $z$.  

As both Figure~\ref{plate} and Table~\ref{table} demonstrate, there is
an excellent agreement between SDSS and \hi\ redshifts (the
largest discrepancy is $\Delta z =0.0004$ for AGC 252580 and AGC
224321. However, the difference drops to $\Delta z =0.0001$ for
AGC 224321 when the SDSS {\em emission line redshift}\footnote{
Not available for AGC 252580.}
is used instead. The emission line redshift of AGC 212941,
0.2238, is also in slightly better agreement with the \hi\ one). 
The importance of having accurate SDSS redshifts is
exemplified by AGC 242147, the highest redshift detection, whose
spectrum shows strong RFI at 1137 MHz.
Because of the long integration required and because in the first hour
or two of observation one can see the RFI but not the galaxy's line
emission, it is essential to know in advance precisely where the
\hi\ signal is expected to appear.

In \S~\ref{s_survey} we noted that, at fixed redshift and \hi\ flux,
galaxies that appear to be slower rotators (i.e., either intrinsically
slower or just viewed more face-on) are significantly easier to detect.
This point is clearly illustrated by the two galaxies shown in the 
second row of Fig.~\ref{plate}, AGC 262033 and AGC 224321, which have 
among the largest and the smallest {\em observed} velocity widths in this
sample, respectively. These are also the objects that required the
longest and the shortest integrations, respectively, in spite of
having almost the same redshift and \hi\ flux.

The rotational velocities deprojected to edge-on view 
(i.e., 0.5 \whicor $/$sin~$i$) vary between \about 180 and 320 \kms\
for this sample.  % 179 and 315 km/s, to be precise.
These values are consistent with those typical of local
disk galaxies with similar rest-frame \iband\ luminosities, as shown for a
large set of \Ha\ rotation curves by \citet[][Fig. 1]{templates}, also
taking into account that optical velocity widths tend to be systematically
smaller than 21 cm ones (a bias understandable in terms of rotation
curve shapes and relative extent of \hi\ vs. \Ha\ emission; see, e.g., 
\citealt{widths}).

We computed gas mass fractions, \Mhi/\Mstar, for the galaxies in our sample. 
Stellar masses, \Mstar, were estimated from
SDSS $i$-band model magnitudes and $g-i$ colors, using
\citet{bell03} stellar mass-to-light vs. color correlations\footnote{
As \citet{bell03} note, their choice of stellar initial mass function
yields ``the maximum possible stellar $M/L$ ratio''. Thus, at given
luminosity, we might overestimate the stellar masses and therefore
underestimate the gas mass fractions.
}. 
We corrected the SDSS magnitudes for Galactic attenuation first, and
applied \citet[][hereafter S07]{s07} K- and internal extinction
corrections (the extinction in SDSS $g, i$ bands was obtained from that in 
S07 Cousins $I$ assuming a standard attenuation curve of the
form $\tau_{\lambda} \propto \lambda^{-0.7}$). The
objects detected at Arecibo are very massive, gas-rich galaxies, with
gas mass fractions \Mhi/\Mstar\ of \about 10$-$30\%, reflective of
the deliberate selection of galaxies on the basis of their
potentially high \hi\ content. % 9-25%, to be precise.

%-------------------------------------------------------------------------------
\subsection{Comparison with $z=0$ field galaxies}\label{ss_res}
%-------------------------------------------------------------------------------

\begin{figure}
\epsscale{0.7}
\plotone{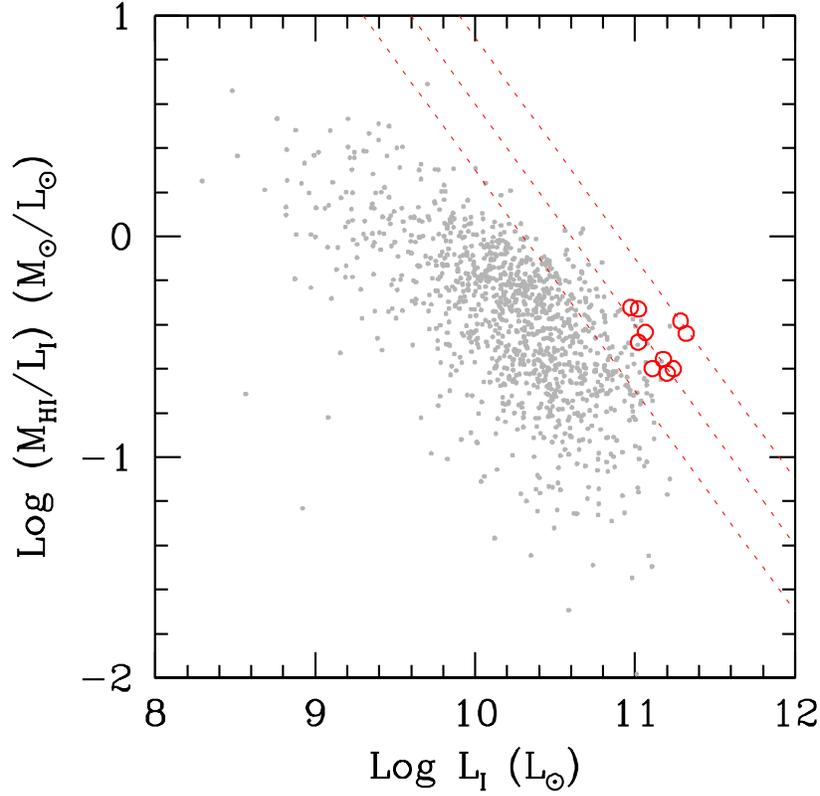}
\caption{\hi\ mass to \iband\ luminosity ratio as a function of
\iband\ luminosity for the Arecibo \hi\ detections (large circles) and
for a comparison field sample at $z=0$ (small symbols). Dotted lines correspond
to constant \hi\ mass, \Mhi $= 2, 4, 8 \times 10^{10}$ \Msun\ (bottom to
top, respectively).
}\label{fig2}
\end{figure}

Since $z\sim 0.2$ corresponds to a non-negligible fraction of the age of
the universe, and in particular, the epoch over which the star formation
density has rapidly declined, it is interesting to investigate if this 
small sample is already suggestive of possible evolutionary effects. 
This exercise requires a representative comparison data set of 
{\em field} galaxies at $z=0$ which we have extracted from the Cornell
\hi\ digital archive (S05). In addition to being a high-quality,
homogeneous compilation of \hi\ parameters for a large number of
galaxies, a subset of the objects in the \hi\ archive
has matching \iband\ photometry and grouping information
from the SFI++ catalog (S07). The cross-match of the S05
\hi\ archive (8850 objects) with the S07 photometry (Table 2: 4053
objects, 2713 of which not listed as group members), after rejection
of galaxies with recessional velocities $cz>10000$ \kms, includes 1081
objects. The redshift cut was applied in order to obtain a clean $z=0$
sample, but the inclusion of the 124 galaxies with $cz>10000$ \kms\ 
(the largest recessional velocity is 17897 \kms) would not change any
of our conclusions.
Since color information is not available for the SFI++
galaxies, we could not estimate stellar masses. Thus we compared 
\hi\ mass to \iband\ luminosity ratios, \MhiLi, for the two samples.
We used standard transformations\footnote{
See {\it http://www.sdss.org/dr6/algorithms/sdssUBVRITransform.html}
(Lupton 2005 equations).
}
to convert SDSS model $r$ magnitudes into Cousins $I$ for our sample,
and rescaled the SFI++ quantities to our adopted Hubble constant.
The resulting \MhiLi\ ratios for our data set (large circles) and for
the $z=0$ reference sample (small symbols) are presented in
Fig.~\ref{fig2}. Dotted lines indicate lines of constant \hi\ mass,
\Mhi $= 2, 4, 8 \times 10^{10}$ \Msun\ (bottom to top, respectively).
The reference sample has a very large scatter, and contains only a
small number of galaxies with rest-frame luminosities comparable with
those of our sample. This limitation makes it very difficult 
to estimate where the
center of the distribution might be, i.e. what is the ``typical'' \hi\
mass of a local {\em field} galaxy with \Li \about $10^{11}$ \Lsun.
The objects in our sample are, on average, very gas-rich, and lie on
or slightly above the upper envelope of the local galaxy distribution.
Naturally, this selection effect arises because we detect only the
gas-richest galaxies at $z=0.2$. Admittedly, galaxies with such large 
\MhiLi\ ratios at these luminosities do appear to be rare in the
local universe. However, since our survey probes a much larger volume
compared to the $z=0$ sample, it is perhaps not surprising to detect
several of these locally rare objects.

Understanding and quantifying evolutionary effects requires the use of
complete samples with well-defined selection criteria. Clearly, this is
not the case for our current survey, which was in part a proof-of-feasibility
experiment. Unfortunately, the $z=0$ reference sample is not
complete in any sense, either. In particular, it is currently very
difficult to find local \hi\ samples with large numbers of galaxies
that are as massive and/or as \hi -rich as our detections. 
Thus, we cannot make any strong statements about the
evolution of the \hi\ mass-to-light ratio based on the results shown
in Fig.~\ref{fig2}.

%-------------------------------------------------------------------------------
\section{Conclusions}\label{s_concl}
%-------------------------------------------------------------------------------

In this Letter, we have shown that carefully-planned
Arecibo observations can deliver high-quality global
\hi\ profiles for galaxies at redshifts up to $z=0.25$ with
integration times of a few hours per object. The detected objects 
are massive disk galaxies with high gas mass fractions
(\about 10$-$30\%), and are apparently rare in the local universe. 
At present,
the small size of our sample, selection effects and the lack of
adequate comparison samples at $z=0$ prevent us from drawing any firm
conclusions about the evolution of the \hi\ content of galaxies
between redshifts 0.25 and 0. 
Larger and well-defined samples of galaxies at $z \sim 0.2$ and higher
are needed in order to make progress in this area. Equally
importantly, fair comparison samples at $z=0$ are {\em also} needed.
The latter might be provided by two large surveys that are ongoing at
Arecibo. The first one is the Arecibo Legacy Fast ALFA (ALFALFA) survey
\citep{g05,g07}, a blind \hi\ survey of the extragalactic sky visible 
from Arecibo, which will make a complete census of \hi-rich galaxies with 
$z \leq 0.060$ and has already detected a significant population of
objects with \hi\ masses $> 10^{10}$ \Msun\ \citep{h08}.
The second one is the GALEX Arecibo SDSS Survey
\citep[GASS;][]{gass}, a recently started program designed to measure 
the \hi\ content of \about 1000 massive galaxies, 
randomly selected from the intersection of the SDSS spectroscopic
survey, GALEX and ALFALFA footprints based on their redshift 
($0.025< z < 0.05$) and stellar mass 
($10.0 < {\rm Log}~[M_{\rm star}/{\rm M}_{\odot}] < 11.5$) only.
ALFALFA and GASS will deliver {\em unbiased} samples of local
\hi -rich and massive galaxies, respectively. In combination with
the future extension of our pilot intermediate redshift field galaxy
program and with complementary \hi\ synthesis mapping programs
of similar galaxies in crowded cluster and group fields, 
it is clearly now feasible  to explore the evolution of
the \hi\ content of disk galaxies over the last several Gyr
for a statistically complete sample.\\

\acknowledgments

BC wishes to thank the Arecibo staff, in particular Ganesan
Rajagopalan and the telescope operators for their assistance, and
Hector Hernandez for scheduling the observations. 
Many special thanks to Phil Perillat for his invaluable help on data
processing and calibration issues. Becky Koopmann and Lisa Wei helped
getting the data reduction pipeline started in the early stages of the survey.
BC also thanks Luca Cortese for useful discussions and Guinevere Kauffmann for
helpful comments on the manuscript. We thank an anonymous referee for
constructive feedback.
This research has made use of the Sloan Digital Sky Survey
archive. Its full acknowledgment can be found at http://www.sdss.org.
We acknowledge the use of the NASA/IPAC Extragalactic 
Database (NED), which is operated by the Jet Propulsion Laboratory, California
Institute of Technology, under contract with NASA.
This work has been partially supported by NSF grants
AST-0307396 and AST-0307661 and by NAIC.\\

{\it Facility:} \facility{Arecibo} (L-wide) \\

%\clearpage

%\clearpage

%\clearpage
%\rotate{
\begin{deluxetable}{cccccccccccc}
\tabletypesize{\scriptsize}
\tablecaption{SDSS and \hi\ Parameters of the Arecibo Detections \label{table}}
\tablewidth{0pt}
\tablehead{
\colhead{}    & \colhead{}      & \colhead{}             & 
\colhead{\Mi} & \colhead{$i$}   & \colhead{$T_{\rm on}$} &
\colhead{}    & \colhead{\whi}  & \colhead{\whicor}      &
\colhead{$F$} & \colhead{} & \colhead{log \Mhi}
\\
\colhead{AGC}   & \colhead{SDSS ID} & \colhead{$z_{\rm SDSS}$}  & 
\colhead{(mag)} & \colhead{(deg)}   & \colhead{(min)}   & 
\colhead{$z$}   & \colhead{(\kms)}  & \colhead{(\kms)}  &
\colhead{(Jy \kms)}  & \colhead{S/N} & \colhead{(\Msun)}
\\
\colhead{(1)}  & \colhead{(2)}  & \colhead{(3)}  & 
\colhead{(4)}  & \colhead{(5)}  & \colhead{(6)}  & 
\colhead{(7)}  & \colhead{(8)}  & \colhead{(9)}  &  
\colhead{(10)} & \colhead{(11)} & \colhead{(12)}
}
\startdata
% agc    sdss_id               z           absM_i   incl    Ton      HI_z        W50       Wcor     flux          s/n  LogMHI
181518 & J082522.13+325953.6 & 0.1711(2) & $-$23.45 & 65 &  112    & 0.1710 & 515$\pm$23 & 417 & 0.26$\pm$0.05 &\phn 8.0 & 10.54 \\
252580 & J151337.28+041921.1 & 0.1754(2) & $-$23.33 & 64 & \phn 84 & 0.1750 & 478$\pm$08 & 384 & 0.32$\pm$0.06 &\phn 8.4 & 10.65 \\
262033 & J162515.41+280530.3 & 0.1870(2) & $-$23.90 & 76 &  260    & 0.1871 & 643$\pm$43 & 518 & 0.24$\pm$0.03 &\phn 9.4 & 10.58 \\
224321 & J120948.14+100822.5 & 0.1879(2) & $-$23.56 & 46 & \phn 80 & 0.1875 & 417$\pm$14 & 328 & 0.26$\pm$0.05 &\phn 8.9 & 10.63 \\
101750 & J003610.70+142246.4 & 0.1909(1) & $-$23.67 & 62 &  156    & 0.1909 & 542$\pm$21 & 432 & 0.19$\pm$0.03 &\phn 8.2 & 10.51 \\
242091 & J140522.72+052814.6 & 0.1954(2) & $-$24.00 & 63 &  156    & 0.1953 & 502$\pm$04 & 397 & 0.25$\pm$0.04 &    10.3 & 10.64 \\
232055 & J134211.36+053211.5 & 0.2031(1) & $-$23.45 & 57 & \phn 92 & 0.2030 & 464$\pm$10 & 362 & 0.26$\pm$0.05 &\phn 8.4 & 10.69 \\
262029 & J160938.00+312958.5 & 0.2188(2) & $-$24.20 & 53 &  136    & 0.2186 & 642$\pm$56 & 503 & 0.34$\pm$0.05 &\phn 9.0 & 10.88 \\
212941 & J111645.15+054210.0 & 0.2239(1) & $-$23.84 & 54 &  180    & 0.2236 & 383$\pm$39 & 289 & 0.18$\pm$0.04 &\phn 7.0 & 10.62 \\
242147 & J142735.69+033434.2 & 0.2454(1) & $-$24.11 & 51 &  176    & 0.2454 & 499$\pm$11 & 377 & 0.28$\pm$0.05 &\phn 8.0 & 10.90 \\
\enddata 
\end{deluxetable}
%}

\end{document}